\documentclass[fleqn,usenatbib]{mnras}
\usepackage[T1]{fontenc}
\usepackage{ae,aecompl}
\usepackage{graphicx}	% Including figure files
\usepackage{amsmath}	% Advanced maths commands
\usepackage{amssymb}	% Extra maths symbols
\title[Cosmic AGN dipoles are much larger than CMB dipole]
{Cosmic dipoles of active galactic nuclei at optical and radio wavelengths display much larger amplitudes than the cosmic microwave background dipole}
\author[A. K. Singal]{Ashok K. Singal\thanks{E-mail: ashokkumar.singal@gmail.com}\\
{Astronomy and Astrophysics Division, Physical Research Laboratory, 
Navrangpura, Ahmedabad - 380009, India}}
\date{Accepted XXX. Received YYY; in original form ZZZ}
\pubyear{2021}
\begin{document}
\label{firstpage}
\pagerange{\pageref{firstpage}--\pageref{lastpage}}
\maketitle

% Abstract of the paper
\begin{abstract}
Sky distributions of large samples of distant active galactic nuclei (AGNs) have shown dipoles significantly larger than the cosmic microwave background (CMB) dipole.
However, a recent Bayesian analysis of the QUAIA sample, comprising 1.3 million quasars, has yielded a dipole that seems to be in tandem with the CMB dipole, in contravention of most previous studies of AGN dipoles. Since the question has large cosmological implications, we investigate the QUAIA quasar sample afresh, by directly computing the dipole from asymmetries observed in the source number counts. We instead find a dipole 3-4 times as large as the CMB dipole though in the same direction. Further, it has been claimed elsewhere that the difference between the CMB dipole and the radio dipole estimated from the NRAO VLA Sky Survey (NVSS), the first large catalogue that showed an AGN dipole about four times larger than the CMB dipole, can be fully accounted for by incorporating the shot-noise and clustering contributions to the total NVSS dipole. A careful reinvestigation of the NVSS dipole, however, shows that the random phenomena like shot noise or clustering cannot account for the actually observed NVSS asymmetries, which show a systematic dipole   pattern over the sky.
\end{abstract}
%--------------------------
%\newpage
\section{INTRODUCTION}
According to the cosmological principle (CP), a comoving observer should find the Universe to be isotropic. However peculiar motion of the observer, a motion with respect to the comoving coordinates of the expanding cosmic fluid, would introduce a dipole anisotropy in some of the observed sky distributions. The CMB, for example, shows a dipole anisotropy which has been interpreted as resulting from a Solar peculiar velocity $370$ km s$^{-1}$ along RA$=168^{\circ}$, Dec$=-7^{\circ}$  (Lineweaver et al. 1996; Hinshaw et al. 2009; Aghanim et al. 2020). 
%\cite{1,2,3}.
%The CMB provides us a reference frame of the Universe at large redshifts $z\sim 10^3$. However, in the last one decade it has been repeatedly seen (Singal 2011; Rubart \& Schwarz 2013; Tiwari et al. 2015; Colin et al. 2017; Bengaly et al. 2018; Singal 2019a,b; Secrest et al. 2021; Siewert et al. 2021; Singal 2021a,b) 
%\cite{4,5,6,7,8,9,10,11,Si21a,Si21b,Si21c} 
Also, dipole anisotropies have been observed in the number counts or sky brightness distributions in the large samples of distant AGNs which have yielded peculiar velocities many times larger than that inferred from the CMB, although mostly in the same direction as the CMB dipole (Singal 2011,19a,b,21a,b,22,23,24; Rubart \& Schwarz 2013; Tiwari et al. 2015; Colin et al. 2017; Bengaly, Maartens \& Santos 2018; Secrest et al. 2021; Siewert, Rubart \& Schwarz 2021; Dam, Lewis \& Brewer 2022; Kothari et al. 2022; Wagenveld, Kl\"ockner \& Schwarz 2023).
%\cite{4,5,6,7,8,9,10,11,Si21a,Si21b}. 
From that it appears that the reference frame of the universe at relatively closer though still at cosmological redshifts ($z\stackrel{>}{_{\sim}} 1$), does not seem to be in conformity with the CMB reference frame at much larger redshifts ($z\stackrel{>}{_{\sim}} 10^3$). 

On the other hand, Mittal, Oayda \& Lewis (2024), from a Bayesian analysis of the Quaia sample of quasars (Storey-Fisher et al. 2023), found a dipole that seems to be in tandem with the CMB dipole. After excising extended Galactic Plane ($|b|<30^\circ$) regions, where the Quaia sample seemed significantly  contaminated by selection effects, Mittal et al. (2024) found evidence that the Quaia quasar dipole is consistent with the CMB dipole, both in terms of the expected amplitude and direction. This result seemed  to lend support to the cosmological principle. 
Since the question at stake, a conformity with the CP, has large cosmological implications, we examine the QUAIA quasar sample afresh, using a more direct approach to compute the dipole from asymmetries observed in the sky distribution of quasar number counts. As we shall show the QUAIA AGN dipole turns out  much larger than the CMB dipole, in tune with the other previously determined  AGN dipoles and thereby  inconsistent with the CP. 

Further, for the NVSS radio catalogue, which was the first large AGN sample that  showed a dipole amplitude about four times as large as the CMB dipole (Singal 2011), it has been claimed recently (Cheng, Chang \& Lidz 2023) that the observed difference between the amplitudes of NVSS and CMB dipoles can be fully accounted for by incorporating the contributions of shot-noise and clustering to the NVSS dipole and it has been the conclusion that the NVSS dipole is consistent with a kinematic origin for the CMB dipole within $\Lambda$CDM. 
However, from a careful examination of the dipole determined from the NVSS data at different flux-density levels and spanning different declination ranges, we show that their observed systematic dipole  patterns over the sky rule out that any random phenomena like shot noise or clustering could account for the actually observed NVSS dipole.
\section{A brief description of the Method}
Following the CP, we expect for a comoving observer the number distribution of quasars to be uniform over the sky.
However, a peculiar motion of the observer along with the Solar system, because of stellar aberration and Doppler boosting, would introduce in the otherwise uniform number density an apparent dipole, along the direction of motion, with an amplitude  
\begin{equation}
\label{eq:7}
{\cal D}=[2+ x (1+\alpha)]\frac {v}{c}\,,
\end{equation}
where $v$ is the peculiar velocity of the  Solar system, $c$ is the velocity of light, $\alpha$ is the spectral index, defined by $S \propto \nu^{-\alpha}$, and $x$ is the index of the integral source counts of extragalactic source population, which follows a power law, $N_0(>S)\propto S^{-x}$ (Ellis \& Baldwin 1984; Crawford 2009). 

%%-------------------------------- ------------
\begin{table*}
\begin{center}
\caption{\label{T1}Dipole estimates for the QUAIA sample for various G-magnitude ($m_{\rm G}$) limits with different galactic latitude (b) cuts.} 
\hskip4pc\vbox{\columnwidth=33pc
%\begin{tabular*}{\textwidth}{@{\extracolsep\fill}ccccccccccccccc}
%\footnotesize
\begin{tabular}{ccccccccccccccc}
\hline 
(1)&(2)&(3)&(4)&(5)&(6)&(7)&(8)\\
Serial&G-magnitude & |b|  & N&\multicolumn{4}{c}{Observed dipole vector and inferred peculiar velocity} \\
\cline{5-8}
%\cline{8-10}
% & ${\cal D}$& ${\cal D}_{\rm h}$  &  & $p_{\rm h}$\\
No.&($m_{\rm G}$)  & limit  & & RA & Dec & ${\cal D}$ &$p$ \\% &  ($10^{-2}$)
&&($^{\circ}$)&&($^{\circ}$)&  ($^{\circ}$)&$10^{-2}$&($370$ km s$^{-1}$)\\
\hline
1&$m_{\rm G}< 20.5$   & $>30$  & 917565 &  $181\pm 14$ &  $+20\pm 13$ & $3.3\pm0.5$ & $4.2\pm0.6$ \\

2&$m_{\rm G}< 20.5$  & $>40$  & 680931 &  $180\pm 15$ &  $+20\pm 14$ & $2.4\pm0.6$ & $3.0\pm0.7$ \\

3&$m_{\rm G}< 20.5$  & $>35$  & 797350 &  $181\pm 14$ &  $+19\pm 13$ & $2.9\pm0.6$ & $3.7\pm0.7$ \\\\
4&$m_{\rm G}< 20.0$  & $>30$  & 530364 &  $179\pm 15$ &  $+17\pm 13$ & $3.3\pm0.5$ & $4.2\pm0.6$ \\

5&$m_{\rm G}< 20.0$ & $>40$  & 395134 &  $175\pm 16$ &  $+16\pm 14$ & $2.3\pm0.6$ & $3.0\pm0.7$ \\

6&$m_{\rm G}< 20.0$ & $>35$  & 461905 &  $178\pm 15$ &  $+14\pm 13$ & $2.9\pm0.6$ & $3.6\pm0.7$ \\\\

7&$20<m_{\rm G}<20.5$  & $>30$  & 387201 &  $184\pm 15$ &  $+24\pm 13$ & $3.3\pm0.5$ & $4.2\pm0.6$ \\

8&$20<m_{\rm G}<20.5$ & $>40$  & 285797 &  $186\pm 16$ &  $+26\pm 14$ & $2.5\pm0.6$ & $3.2\pm0.7$ \\

9&$20<m_{\rm G}<20.5$ & $>35$  & 335445 &  $186\pm 15$ &  $+24\pm 13$ & $3.0\pm0.6$ & $3.8\pm0.7$ \\

%----------------------------------------
\hline
\end{tabular}
%\normalsize
%\end{tabular*}
}
\end{center}
\end{table*}
%%-------------------------------- ------------

A vector sum of the angular position vectors of all sources in a sample having a full sky-coverage gives the direction of the dipole as well as provides a measure of the dipole amplitude. 
This is called the dipole vector method (Crawford 2009; Singal 2011,24). 
We can investigate the dipole using another way, known as hemisphere method, by computing projection of the dipole in various directions of the sky. For this, we first divide the sky into $10^\circ \times 10^\circ$ pixels, creating a grid of $422$ cells covering the whole sky. 
Then taking the centre of each of these 422 cells, we divide the sky in two equal hemispheres, $\Sigma_1$ and $\Sigma_2$, with $\Sigma_1$ centred on the chosen cell and $\Sigma_2$ as the opposite hemisphere and count the number of sources $N_1, N_2$ within $\Sigma_1, \Sigma_2$ respectively in our sample. Then $2(N_1-N_2)/(N_1+N_2)$ gives the component of the dipole in the direction of the chosen cell (Singal 2023,24). A contour map of thus computed dipole components yields the direction and amplitude for the dipole. The pole direction is expected to be in vicinity of the highest contour value.
We expect a  $\cos\psi$ dipole pattern with respect to the true dipole direction in sky, with a minimum of a chi-square fit in that direction, which should agree with the corresponding value  derived directly from the dipole vector method in case of a genuine dipole. 
%--------------------------------------------

For a partial sky-coverage with symmetric cuts in diametrically opposite regions, e.g. $|b|<30^\circ$, which affect the forward and backward measurements 
identically, there is no effect on the direction inferred for the dipole (Ellis \& Baldwin 1984, Singal 2011), except that position errors may be higher because of the smaller sample. The dipole amplitude may need, however, to be corrected in such cases by factors of order unity (Singal 2011,24).

%-------------------------------------------------------
\begin{figure}
\includegraphics[width=\columnwidth]{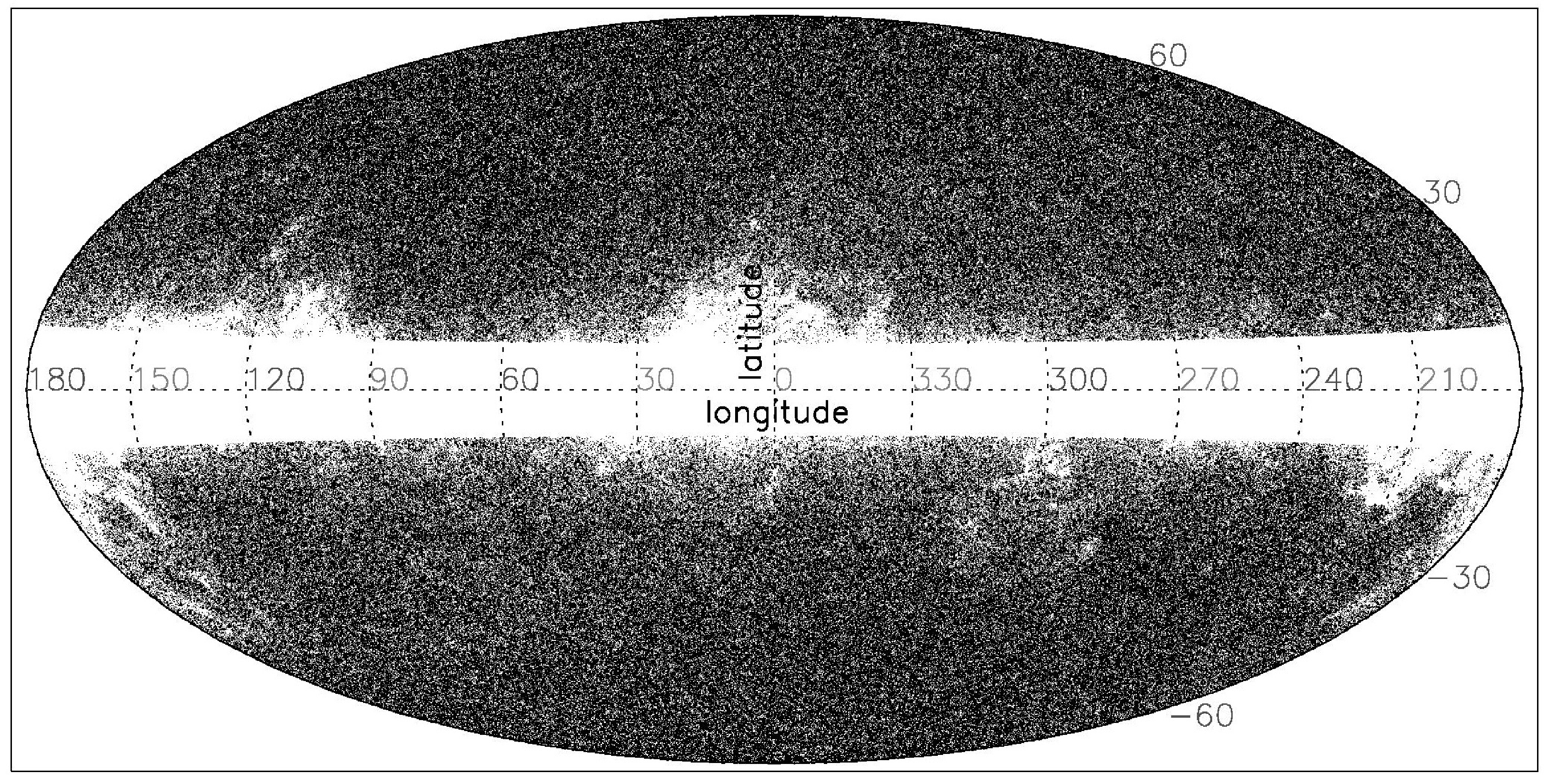}
\caption{The sky distribution of 1.3 million quasars in the QUAIA $m_{\rm G}<20.5$, $|b|>10^\circ$  sample, in~the Hammer--Aitoff equal-area projection map, plotted in galactic coordinates. 
\label{F1}
}
\end{figure}
%%----------------------------------------------------------------

\begin{figure}
%\centering
\includegraphics[width=\columnwidth]{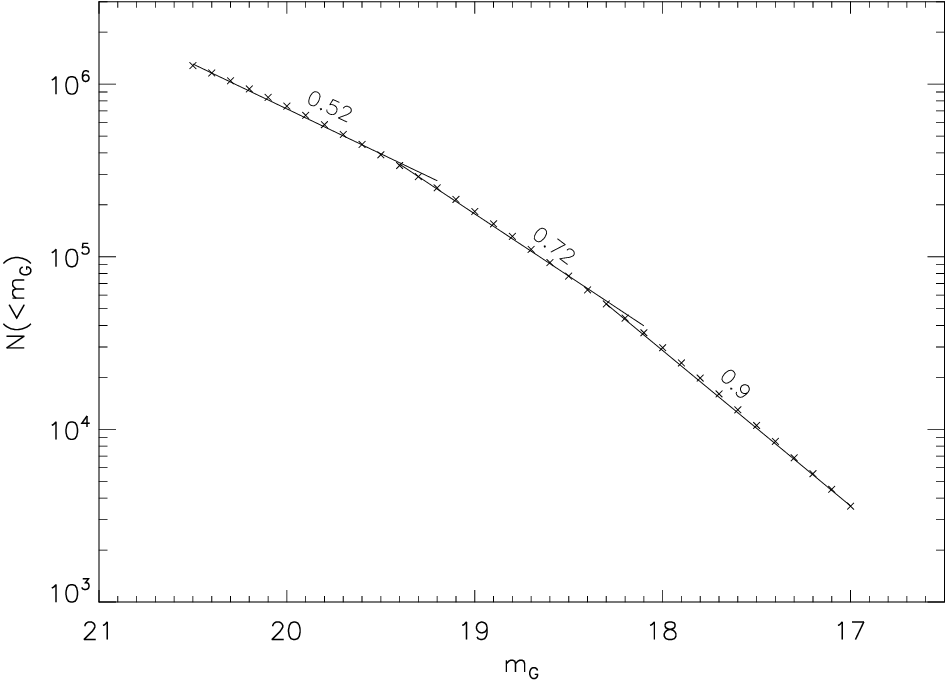}
\caption{A plot of the integrated source counts $N(< m_{\rm G})$ against $m_{\rm G}$ for the QUAIA sample, showing the power law behavior of the source~counts. From  piece-wise straight line fits to data in different $m_{\rm G}$ ranges, the slope appears to steepen for stronger sources, as shown by continuous lines with the slope values written above.
\label{F2}
}
\end{figure}
%\unskip

%--------------------------------------------
%-----------------------------------------------------------------
\section{Dipole seen in the sky distribution of QUAIA quasars}
%--------------------------------------------

%\section{QUAIA sample of quasars}
QUAIA, the Gaia-unWISE Quasar Catalog, is a publicly available, an all-sky spectroscopic quasar sample, which may be a highly competitive sample for cosmological large-scale structure analyses (Storey-Fisher et al. 2023). The sample is drawn from the 6,649,162 quasar candidates, identified by the Gaia mission (Gaia Collaboration 2016), released in Gaia DR3 (Gaia Collaboration 2023a,b). Further, all Gaia
quasars have been cross-matched with those from the Wide-field Infrared Survey Explorer (WISE; Wright et al. 2010), to also provide photometric information in the W1 and W2 infrared bands. 
The QUAIA catalogue is available in two versions with different Gaia G-band magnitude limits, the full  $m_{\rm G}< 20.5$ version containing 1,295,502 quasars, and a reduced but cleaner version with 755,850 quasars, which is just a subset of the larger catalogue with an additional magnitude cut, $m_{\rm G}< 20.0$. 

Figure~\ref{F1} shows a Hammer--Aitoff equal-area projection plot in galactic co-ordinates of the QUAIA quasars with $m_{\rm G}< 20.5$, $|b|>10^\circ$. 
According to Storey-Fisher et al. (2023), the QUAIA sample has significantly lower accuracy around the Galactic Plane, as can be seen from Fig.~\ref{F1} here. To mitigate its effects on the determined dipole, we restrict the Galactic latitude in our sample to $|b|>30^\circ$. As noted by Mittal et al. (2024), the $|b|<30^\circ$ mask should cover much of the problematic regions, but to counter the possibility that some issues of non-uniformity at the edge of the mask (Fig.~\ref{F1}) may still seep into our sample, following Mittal et al. (2024) we also employ $|b|<40^\circ$ mask. Apart from that, *

%The sky distribution of quasars in galactic coordinates is displayed in Fig.~1, where we have indicated the positions of the CMB pole ($\odot$).

To determine the power law index $x$ for our sample of quasars, we have made a plot of the integrated source counts $N(< m_{\rm G})$ against $m_{\rm G}$, in~Fig.~\ref{F2}, which shows a power law behavior of the integrated source counts, with~a slope that varies between 0.52 and 0.9. As emphasized by Singal (2023), the relevant values of $x$ and $\alpha$ are the ones in the vicinity of the lower  threshold flux density and are determined empirically from actual observations. 
Since the index in the $N(< m_{\rm G})$ plots (Fig.~\ref{F2}) for the weakest sources in our two samples, $m_{\rm G}=20.5$ and $m_{\rm G}=20.5$, is 0.52, accordingly the index of integral source counts in our sample is $x=2.5 \times 0.52= 1.3$. For~the spectral index, we have taken 
$\alpha\approx 2.4$ for both samples (Mittal et al. 2024). The~peculiar speed, accordingly, is determined in our sample from, $v\approx c {\cal D} /6.4$.
%$\alpha\approx 2.37$ for the  $m_{\rm G}< 20.5$ sample and $\alpha\approx 2.44$ for the  $m_{\rm G}< 20.0$ sample (Mittal et al. 2024). The~peculiar speed, accordingly, is determined in our sample from, $v\approx c {\cal D} /(6.38)$ for the  $m_{\rm G}< 20.5$ sample and $v\approx c {\cal D} /(6.47)$ for the  $m_{\rm G}< 20.0$ sample. 
For convenience of a comparison with the CMB dipole, we use a parameter $p$ to express the peculiar velocity $v$, in units of the CMB value, so that $v=p \times 370$ km s$^{-1}$. It means a peculiar velocity equal to the CMB value is represented by $p=1$. Of course, because of large gaps because of the Galactic Plane masks, we apply  appropriate corrections, which are of the order of unity (Singal 2024), in the dipole amplitudes. 
%--------------------------------------------
\begin{figure}
\includegraphics[width=\linewidth]{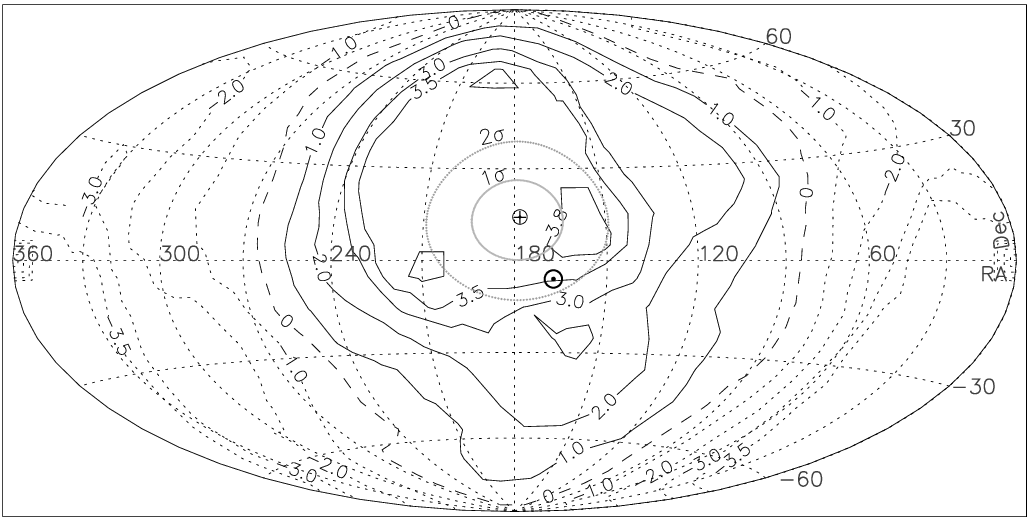}
\caption{A contour map of the dipole amplitudes, estimated for various directions in the sky, from the QUAIA $m_{\rm G}<20.0$, $|b|>35^\circ$ data. The contour values depict components of the estimated peculiar velocity, in units of the CMB value of 370 km s$^{-1}$, in various directions of the sky. The~horizontal and vertical axes denote RA and Dec in degrees. Positive component values are shown by continuous contour lines, while the negative component values are shown by dotted lines. The~dashed curve represents zero amplitude of the dipole component. 
%The symbol $\oplus$ indicates the derived pole position for the QUAIA sample, along with   $1\sigma$ and $2\sigma$ error ellipses around it. 
The~optimum pole direction is expected to be in vicinity of the highest contour value that seems to lie in close proximity of the CMB pole position, indicated by symbol $\odot$. 
The symbol $\oplus$ indicates the optimum pole position where a 3D $\cos\psi$ fit yields a minimum chi-square, while $1\sigma$ and $2\sigma$ errors are indicated by grey ellipses around it.
%The symbol $\odot$ indicates the CMB pole position. 
\label{F3}
} 
\end{figure}
%--------------------------------------------

%%--------------------------------------------
\begin{table*}
\begin{center}
\caption{\label{T2}Velocity vector from number counts for the NVSS dataset in different declination ranges with $|b|>10^\circ$.}
\hskip4pc\vbox{\columnwidth=33pc
\begin{tabular}{ccccccccccccccc}
\hline 
(1)&(2)&(3)&(4)&(5)&(6)&(7)&(8)\\
Serial&Flux-density & Declination  & $N$ 
&\multicolumn{4}{c}{Peculiar velocity vector estimate} \\
\cline{5-8}
No.&$S$ &range & &  RA & Dec & ${\cal D}$  &   $p$\\
%& ${\cal D}_{\rm h}$  &  & $p_{\rm h}$\\
&(mJy)&($^{\circ}$)&& ($^{\circ}$)&  ($^{\circ}$) &  ($10^{-2}$) & ($370$ km s$^{-1}$)&  \\%($10^{-2}$)& ($370$ km s$^{-1}$) \\ 
\hline
%-----------------------------------------
1 & $\geq 18$ & $|{\delta}|\leq40$ & 252842 &  $157\pm 10$ &  $+06\pm 10$ & $1.7\pm 0.3$ & $4.0\pm0.8$ \\%& $2.5\pm 0.4$ & $5.4\pm0.9$ \\
2 & $\geq 18$ & $20<|{\delta}|\leq40$ & 116596 &  $165\pm 15$ &  $+07\pm 14$ & $2.0\pm 0.5$ & $4.5\pm1.1$ \\%& $2.3\pm 0.6$ & $5.1\pm1.3$ \\
3 & $\geq 18$ & $|{\delta}|\leq20$ & 136246 &  $149\pm 14$ &  $+05\pm 13$ & $1.6\pm 0.4$ & $3.7\pm1.0$ \\%& $2.3\pm 0.6$ & $5.1\pm1.3$ \\
%-----------------------------------------
\hline
\end{tabular}
}
\end{center}
\end{table*}
%--------------------------------------------

Table~\ref{T1} gives the dipole estimates for both QUAIA samples $m_{\rm G}< 20.5$ and $m_{\rm G}< 20.0$, with different Galactic Plane cuts ($|b|>30^\circ, |b|>40^\circ$). We have also made an intermediate cut $|b|>35^\circ$. For either of the $m_{\rm G}< 20.5$ or $m_{\rm G}< 20.0$ QUAIA samples, the dipole amplitude is significantly higher (about three to four times) than the CMB dipole. We also used the QUAIA   sample with $20<m_{\rm G}<20.5$. As can be seen from Table~\ref{T1}, a comparison of the results from rows 7-9 with those in rows 4-6 shows consistent values for the dipole estimates, both on position and amplitude, from QUAIA quasars in the $20<m_{\rm G}<20.5$ sample vs. in the $m_{\rm G}<20.0$ sample. Now these two quasar samples are independent of each other, with not even a single source overlapping. A consistency of the inferred dipoles in these two statistically-independent samples gives us confidence in the dipole values. 

We do not know Why the Bayesian analysis by Mittal et al. (2024) had yielded a different results for the dipole. We have used the version v0.1.0 of Quaia, the same as used by Mittal et al. (2024). We 
did not incorporate the selection function provided by (Storey-Fisher et al. 2023). The reason being that their procedure seems to  obliterate
any dipole asymmetries across the sky as according to Storey-Fisher et al. : "In
the case of no selection effects (and under the assumption of
isotropy), we would expect each pixel to contain roughly the same
number of sources." Later they mention: "We take the mean predicted
number of quasars in these clean pixels, normalize the predicted
source numbers by this mean, and then fix pixels with normalized
values greater than 1 to exactly 1." In our opinion the dipoles will
get suppressed in this procedure and that might be the reason why results of Mittal et al. (2024) differ from ours.

Figure~\ref{F3} shows a contour map of our computed dipole components. The dipole direction for the corresponding QUAIA sample (Table~\ref{T2}), where a 3D $\cos\psi$ fit yields a minimum chi-square value (Singal 2023,24), lies on the maxima of the contour map, which lies very close to the CMB dipole position on sky. However, the maxima of the contour level, representing the dipole amplitude, is at least a factor of three higher than the CMB dipole, in tune with the other previously determined  AGN dipoles, which are inconsistent with the CP. 

\section{The case of NVSS dipole}
The NRAO VLA Sky Survey (NVSS), comprising 1.8 million radio sources with a flux-density limit >3 mJy at 1.4 GHz (Condon et al. 1998), was the first large catalogue that showed in the sky distribution of distant AGNs a dipole asymmetry about four times larger (Singal 2011) than the CMB dipole asymmetry, though the direction of the AGN dipole turned out strangely in the same direction as the CMB dipole. 
A large dipole in the NVSS data has subsequently been confirmed (Singal 2019a,23,24; Rubart \& Schwarz 2013; Tiwari et al. 2015; Colin et al. 2017; Bengaly et al. 2018; Siewert et al. 2021; Secrest et al. 2022; Wagenveld et al. 2023). 
%Curiously, though the amplitudes of AGN dipoles were almost always much larger than of the CMB dipole, the direction turned out in a narrow range around the CMB dipole direction. From CP one expects the Universe at large scales to be homogeneous and isotropic, in such a scenario it is accordingly believed that any anisotropies seen may be resulting from observer's peculiar motion along with the Solar system.However, such discrepancies between the AGN and CMB dipoles cannot occur due to a genuine peculiar motion of the Solar system and raises doubt whether the CP really holds true. 
However, it has been recently claimed that the large amplitude in the NVSS data, could be resulting from the shot noise in the observed numbers of sources and the occurrence of clustering (Cheng et al. 2023), and that the NVSS dipole could be in tune with the CMB dipole, both in direction and amplitude, as expected according to CP. 
%------------------------------------------
\begin{figure}
\includegraphics[width=\linewidth]{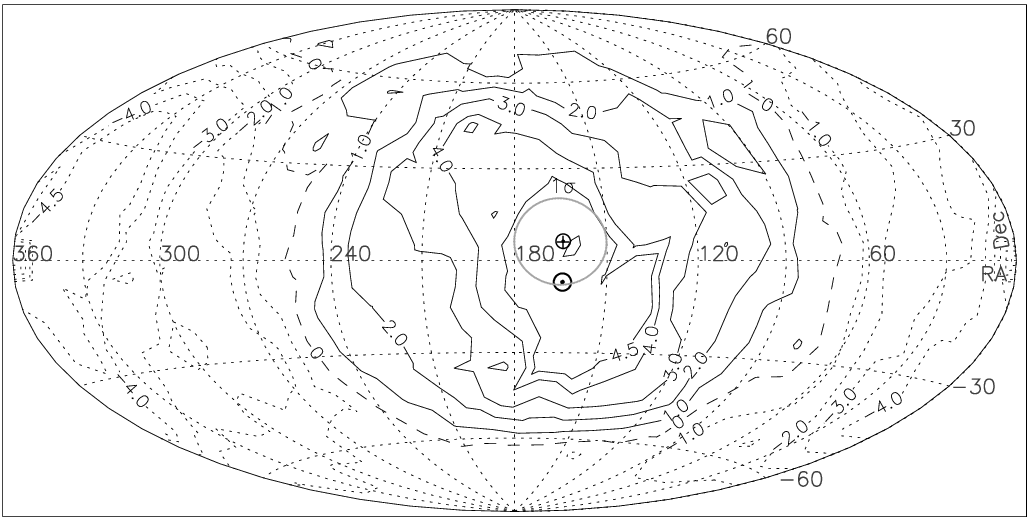}
\caption{A contour map of the dipole amplitudes, estimated for various directions in the sky, for the NVSS data with $S\geq 18$, in the range $20<|{\delta}|\leq40$. The contour values depict components of the estimated peculiar velocity, in units of the CMB value of 370 km s$^{-1}$, in various directions of the sky. The~horizontal and vertical axes denote RA and Dec in degrees. Positive component values are shown by continuous contour lines, while the negative component values are shown by dotted lines. The~dashed curve represents zero amplitude of the dipole component. 
%The symbol $\oplus$ indicates the derived pole position for the QUAIA sample, along with   $1\sigma$ and $2\sigma$ error ellipses around it. 
The~optimum pole direction is expected to be in vicinity of the highest contour value that seems to lie in close proximity of the CMB pole position, indicated by symbol $\odot$. 
The symbol $\oplus$ indicates the optimum pole position where a 3-D $\cos\psi$ fit yields a minimum chi-square, while $1\sigma$ and $2\sigma$ errors are indicated by grey ellipses around it.
%The symbol $\odot$ indicates the CMB pole position. 
\label{F4}
} 
\end{figure}
%--------------------------------------------
%%--------------------------------------------
\begin{figure}
\includegraphics[width=\linewidth]{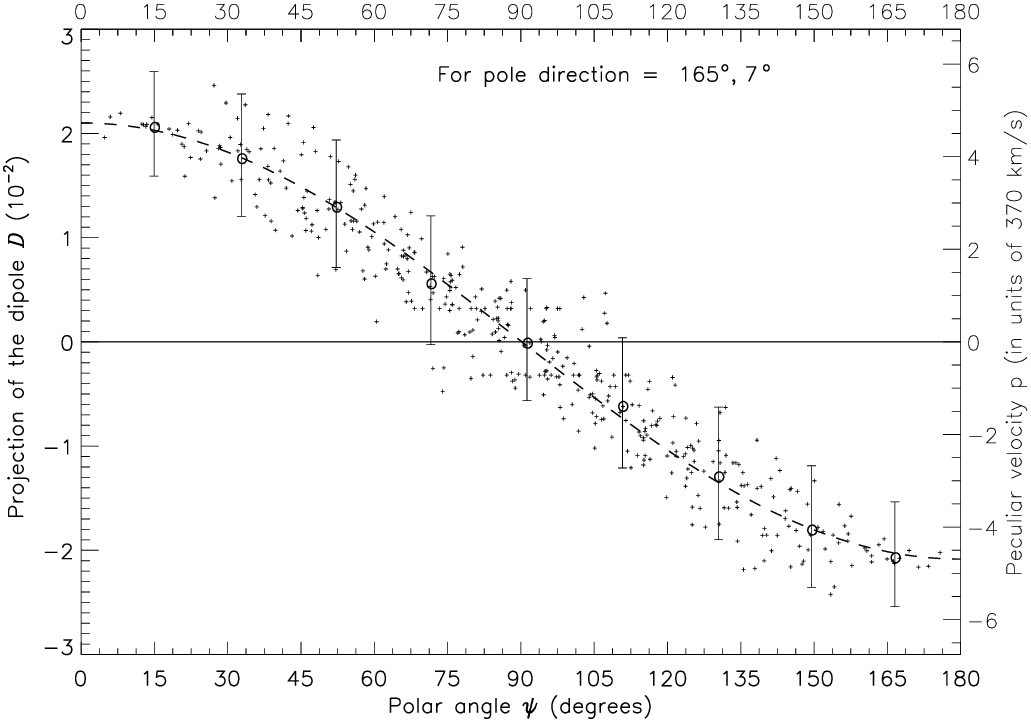}
\caption{The dipole component ${\cal D}$, computed for various polar angles %(a) 
with respect to the dipole direction, RA$=165^{\circ}$, Dec$=7^{\circ}$ (derived by the dipole vector method, Table~\ref{T2}) for the NVSS sample in the range $20<|{\delta}|\leq40$. 
The corresponding component values of peculiar velocity $p$ (in units of CMB value 370 km s$^{-1}$) are shown on the right hand vertical scale.
Plotted circles (o) with error bars show values for bin averages of the dipole components, obtained for various $20^{\circ}$ wide slices of the sky in polar angle, while the dashed line shows a least square fit of $\cos \psi$ to the bin average values. 
\label{F5}
%Variation of the peculiar velocity component $p$ (in units of CMB value 370 km s$^{-1}$), computed for various polar angles %(a) 
%with respect to the dipole direction, RA$=165^{\circ}$, Dec$=7^{\circ}$, derived by the dipole vector method for the NVSS sample (Table~\ref{T2}) in the range $20<|{\delta}|\leq40$. 
%The corresponding peculiar velocity values of the Solar system in units of $10^3$ km s$^{-1}$ are shown on the right hand vertical scale.
%Plotted circles (o) with error bars show 
%values for bin averages of the peculiar velocity components, obtained for various $20^{\circ}$ wide slices of the sky in polar angle,  
%while the dashed line 
%shows a least square fit of $\cos \psi$ to the bin average values.\label{F5}
}
\end{figure}
%%--------------------------------------------
%%--------------------------------------------
\begin{figure}
\includegraphics[width=\columnwidth]{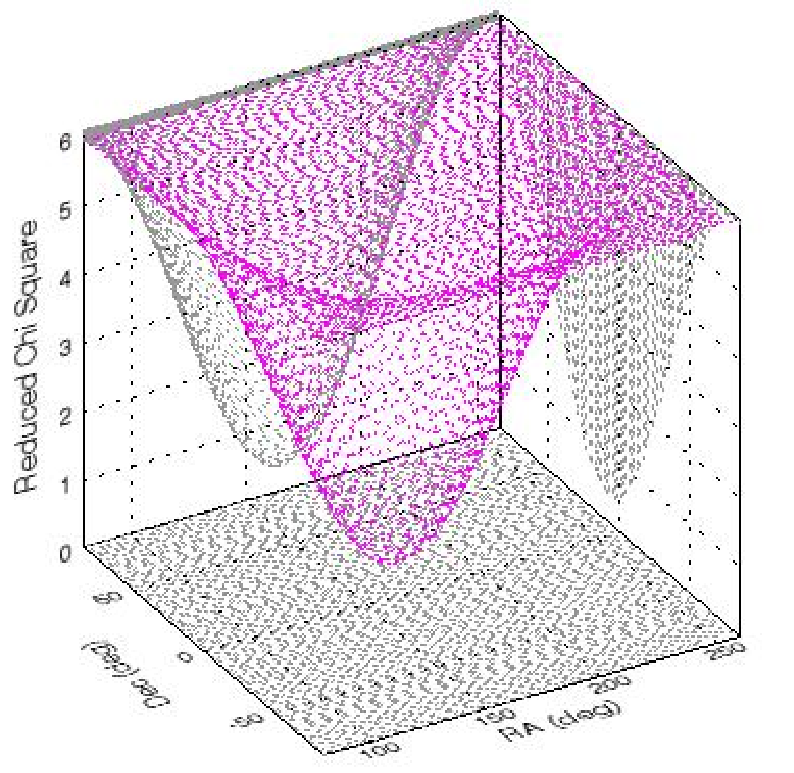}
\caption{A 3-D plot of the reduced chi-square ($\chi^2_\nu$) values (in violet colour), from the cosfit routine for various trial directions for the NVSS data with $S\geq 18$, in the range $20<|{\delta}|\leq40$. The horizontal plane shows the direction in sky as RA and Dec in degrees. The position (RA and Dec) of the minimum of the reduced chi-square is determined at RA$=165^{\circ}$, and Dec$=07^{\circ}$ from the 2-D projections, shown in light grey.}
\label{F6}
\end{figure}
%--------------------------------------------
%%--------------------------------------------

It should be noted here that the NVSS dipole has been determined at various flux-density levels, where in all cases the dipole yielded similar values, with amplitude about four times the CMB dipole and direction similar to that of the CMB dipole (Singal 2011). Such would not happen if the inferred larger amplitude of the dipole were mainly due to the random effects like shot noise. Moreover the errors for the dipole amplitudes in Singal (2011) had already taken into account the shot noise in the number counts. These  estimated errors were much smaller than the computed amplitude values. Also, in order to see if there were any effects of any local clustering in the direction of the Virgo Supercluster, the dipole amplitude was determined after making cuts in NVSS data in progressively increasing Supergalactic latitude values, but no significant changes were seen in the computed dipole amplitude. From that it was concluded that the observed amplitude of the NVSS dipole being larger than the CMB by a factor of four is not resulting from a local clustering.

Further, in Singal (2024), the NVSS dipole was computed not only at different flux-density levels, but now for any  given flux-density range, say for $S>18$mJy, the dipole was determined for various sub-samples with different declination ranges, though with considerable overlaps. Again, different sub-samples gave  consistent values for the direction and amplitude of the NVSS dipole. Now the shot noise as well as  clustering in different sub-samples covering different declination regions cannot be the same so as to yield the same dipole in all cases. 

Here one should note that the shot noise as well as clustering will cause random fluctuations in the dipole amplitude as well as in its direction. It would be a rather contrived scenario if in all cases the direction of the dipole affected by shot noise as well as clustering were to yield much larger dipole but always toward the CMB dipole direction. In the case of NVSS the observations have yielded the same large value of the dipole amplitude (about four times the CMB dipole) in the same direction, even when the samples were used with different lower  cutoff levels (Singal 2011,19a,23). Such would not occur except in a very much contrived case. 

%--------------------------------------------
Since the declination ranges chosen in Singal (2024) had substantial overlaps, with a number of common sources, we attempt here declination ranges with no overlap. Table~\ref{T2} shows the dipole position and amplitude for the $S>18$mJy sample for three declination ranges, ${|\delta}|\leq 40^\circ$, $20^\circ<{|\delta}|\leq40^\circ$ and ${|\delta}|\leq 20^\circ$. 
In fact, the sky-coverage limit $20^\circ<{|\delta}|\leq40^\circ$ in the 2nd row has no overlap with the sky-coverage limit, ${|\delta}|\leq 20^\circ$ in the 3rd  row of Table~\ref{T2}, where, with not even a single common source and thus statistically independent shot noise as well as clustering, if any, the dipole in each case turns out to be not only very similar in amplitude, but also the  direction turns out so close to the CMB dipole direction in either case. 

Figure~\ref{F4} shows a contour map of the computed dipole for various sky points. For this, as in Fig~\ref{F3}, the sky was divided into $10^\circ \times 10^\circ$ pixels, creating a grid of $422$ cells covering the whole sky and the component of the dipole in the direction of each cell was computed. In order to compute the expected dipole amplitude, we used $\alpha=0.65$ and $x=0.95$ for the NVSS data (Singal 2024). The contours appear to have a systematic angular pattern over the sky with a maxima close to the CMB dipole position in sky and the amplitude close to four times the CMB dipole amplitude. If the inferred dipole amplitude $p \sim 4$ were instead due to something like a shot noise or clustering, both random phenomena, then we should be seeing random variations of that order in amplitude distribution over the sky which we do not see. It is clear that the dipole amplitude of $p \sim 4$ is not resulting from shot noise or clustering and is genuinely present even if otherwise we may not be sure of its cause of origin, whether Solar peculiar motion or some other as yet unknown, but may be some intrinsic effect.
%--------------------------------------------

The shot noise or clustering might introduce some random fluctuations in the computed dipole, both in direction and amplitude, but then one would not expect a systematic $\cos \psi$ variation of the dipole pattern with respect to the maximum value. To investigate that, we have plotted in Fig.~(\ref{F5}) the dipole component ${\cal D}$, for various sky points, as a scatter plot for various polar angles, measured with respect to the dipole direction,  RA$=165^{\circ}$, Dec$=7^{\circ}$, derived from dipole vector method, (Table~\ref{T2}, for the NVSS sample in the range $20<|{\delta}|\leq40$. 
We also computed bin averages of the dipole component ${\cal D}$ in $20^{\circ}$ wide slices of the sky by divided the sky into bins of $20^\circ$ width in polar angle about the above pole position, viz. RA$=165^{\circ}$, Dec$=7^{\circ}$. The scatter plot and their various bin-average values clearly show the expected $\cos\psi$ behaviour with a maximum value of $p=4.5$ (Fig.~\ref{F5}). 
A least square fit of $\cos \psi$ to the bin average values (Fig.~(\ref{F5})) shows that the computed ${\cal D}$ values for various sky points at polar angles ($\psi$) do follow a systematic $\cos\psi$ dependence. 
The corresponding component values of peculiar velocity $p$ (in units of CMB value 370 km s$^{-1}$) are shown on the right hand vertical scale.
%%--------------------------------------------

In case of a genuine dipole, one expects a  $\cos\psi$ dipole pattern with respect to the true dipole direction in sky, with a minimum of a  chi-square fit in that direction. 
To check that, we made a 3-D $\cos\psi$ fit for each of the $n=422$ positions for the remaining $n-1$ $p$ values, and computed the chi-square value for each of these $n$ fits. 
A reduced Chi-squared ($\chi^2_\nu$) values for the 3-D  cos fits  made to the dipole amplitudes estimated for various trial dipole directions across the sky, shows a clear minimum  (Fig.~(\ref{F6})), from which we infer the direction of the observer's peculiar velocity as RA = 165$^{\circ}$, and Dec = 7$^{\circ}$, which agrees very well with the corresponding value  derived directly from the dipole vector method (Table~2, $20<|{\delta}|\leq40$).
%--------------------------------------------

\section{Conclusions}
It was shown that the QUAIA quasar dipole, which had earlier been shown through a Bayesian analysis to be consistent with the CMB dipole, shows from a direct determination to be about three to four times larger in amplitude. Also the NVSS dipole, which was recently claimed to be affected by random shot noise and clustering, has a systematic dipole pattern in sky which cannot be due to som ransom precesses.  The dipoles determined from the asymmetries seen in the angular distribution of AGNs in sky appear almost always to be of significantly larger amplitudes than the CMB dipole, although their directions seem to overlap  within statistical uncertainties. Such large differences in the amplitudes of the AGN and CMB dipoles but non-random orientations in sky cannot be due to some random process like shot noise or clustering in the AGN number distributions, instead it  suggests a preferred direction in the Universe, which contrary to the conventional wisdom is not due to a peculiar motion of the Solar system, raises doubts about the CP, the basic foundation of the standard model in modern cosmology.
%%%%%%%%%%%%%%%%%%%%%%%%%%%%%%%%%%%%%%%%%%
%\newpage
\section*{Declarations}
The author has no conflicts of interest/competing interests to declare that are relevant to the content of this article. No funds, grants, or other support of any kind was received from anywhere for this research.
%%%%%%%%%%%%%%%%%%%%%%%%%%%%%%%%%%%%%%%%%%
%--------------------------------------------
\section*{Data Availability}
The data used in this article is freely available at DOI 10.5281/zenodo.8060755. 
%--------------------------------------------
%--------------------------------------------

\end{document}